# System Level Analysis of the Bluetooth standard

*Massimo Conti, Daniele Moretti*

Università Politecnica delle Marche, via Brecce Bianche, I-60131, Ancona, Italy

**Abstract**

*The SystemC modules of the Link Manager Layer and Baseband Layer have been designed in this work at behavioral level to analyze the performances of the Bluetooth standard. In particular the probability of the creation of a piconet in presence of noise in the channel and the power reduction using the sniff and hold mode have been investigated.*

## 1. The Bluetooth Standard

Bluetooth is an emerging standard for short distance wireless communications developed by the Bluetooth Special Interest Group (SIG) [1].

Bluetooth is a short-range (10–100m) wireless link technology aimed at replacing cables that connect phones, laptops, PDAs, and other portable devices.

The Bluetooth standard operates at 2.4 GHz in the ISM band (Industrial Scientific Medicine) with GSFK modulation (Gaussian Frequency Shift Keying). The FHSS (Frequency Hopping Spread Spectrum) technique is used to reduce the effect of radio frequency interferences on transmission quality. The Bluetooth devices sharing the same channel form a network called piconet, with a single unit acting as a master, the other units acting as slaves. Up to eight devices constitute a piconet, with a master device coordinating access by a polling scheme.

The channel is represented by a pseudo-random hopping sequence in the 79 RF channels of 1-MHz width. The raw data rate is 1 Mbit/s. The hopping sequence is unique for the piconet and is determined by the Bluetooth address of the master. The channel is divided into time slots, where each slot corresponds to an RF hop frequency. Consecutive hops correspond to different RF frequencies. The nominal hop rate is 1600 hops/s.

A time division multiplexing (TDD) technique divides the channel into 625 μsecs slots and, with a 1-Mbit/s symbol rate, a slot can carry up to 625 bits.

The standard defines two types of link between master and slaves: Synchronous Connection-Oriented (SCO) link, and Asynchronous Connection-Less (ACL) link.

Many low cost Bluetooth devices are now available and many papers have been presented for the hardware implementation [2 and reference therein]. The Bluetooth protocol is complex and many research is in progress for choice of the parameters of the standard in order to optimize the efficiency of the standard itself.

For example, Cordeiro [3] developed a model implementing the basic functionality of the Baseband, LMP and L2CAP layers to evaluate the impact of interference on the throughput. In [4,5] a theorical analysis of packet error rate of a Bluetooth piconet in proximity with other Bleutooth piconets and 802.11 WLANs is presented.

## 2. Bluetooth Architecture in SystemC

The Bluetooth Stack, in Fig.1, goes from the high level Application layer to the low level radio frequency layer, the lower layers has been modeled in this work using the SystemC language. SystemC is an emerging standard modeling platform based on C++ that supports design abstraction at the RTL, behavioral and system level [6].

Design methods for integrated circuits based on hardware description languages (such as VHDL or Verilog), when applied to these systems, lead to long simulation time and force the designer to introduce useless details about the design; these methods are not suitable for SoC design that requires a high abstraction level and efficient IP management and reuse. The use of C/C++ models is profitable to describe complex IP integration and IP behaviors, for which not all internal details are necessary, in order to early confirm the design specifications and to be early delivered to third parties for evaluation.

The SystemC modules of the Link Manager Layer, Baseband Layer and the analog RF Layer have been designed in this work at behavioral level with the aim of:
- create a fast simulation environment of the lower layers of the Bluetooth standard;
- verify the Bluetooth performance in presence of noise;
- define different algorithms at application layer and verify their impact on piconet performances;
- use this behavioural model as a golden model for future design of these Bluetooth layer at RTL level;
- analyze the power dissipation of the digital and RF part in the different phases of the life of a piconet (inquiry, page, active, sniff, park and hold);
- analyze the effect of the use of different type of packets (DH1, DH3, DH5, DM1, DM3, DM5) in the throughput and power dissipation in presence of noise.

Some of these analyses are reported in this paper: the effect of the BER on the creation of the piconet and the choice of different modes in which each slave can operate (active, sniff, park and hold) in order to reduce power dissipation. The modules of the analog RF layer has not been used in this work in order to speed up the simulations. Conversely, in this work the noisy channel has been modelled as a digital module with a single output and multiple inputs, one for each one of the Bluetooth devices, as reported in Fig.2. The channel emulates :
- the effect of noise in the channel with an inversion of the bit in the channel controlled by a random number generator.
- the delay of the modulator and demodulator RF blocks. The Bluetooth protocol takes into account the effect of this delay: the synchronization of the piconet may be lost for an high value of this delay.
- The collision between packets: this case is possible when the piconet is not already created or when two or more piconets coexist. The output of a device that is not transmitting is in high impedance 'Z'. When more than one device is transmitting in the channel the channel resolver forces the signal in input to the receivers to an undefined value 'X' indicating a collision.

Fig. 3 reports the architecture of the Baseband Layer has it has been implemented in SystemC. It consists of the following modules:

**State Machine**: it implements the functionality of the master and of the slave. The main method is sensitive to the Bluetooth clock (1MHz), which is used to send and receive commands to the others modules. It consists of different methods and processes:
- Enable_inquiry: used to enter the state machine in inquiry state;
- Enable_page: used to enter in page state;
- Enable_detach_reset: resets or detach the state machine.
- Enable_inquiry_scan:used to enter in inquiry scan state;
- Enable_page_scan: used to enter in page scan state;
- Enable_sniff_mode: used to enter in sniff mode;
- Enable_hold_mode: used to enter in hold mode;
- Enable_park_mode: used to enter in park mode;

**Transmitter:** it receives the data to be inserted in the packet, and builds up the packet to be send to the RF module;

**Receiver:** it interprets the packet received from the RF module;

**Buffer_tx:** buffer to store data from Link Manager to Baseband;

**Buffer_rx:** buffer to store data from Baseband to Link Manager;

**Piconet:** it takes into account the addresses and the **synchronism of the piconet;**

**Clock:** it generates the clock native;

**Hop-freq:** it generates the correct frequency hopping sequence.

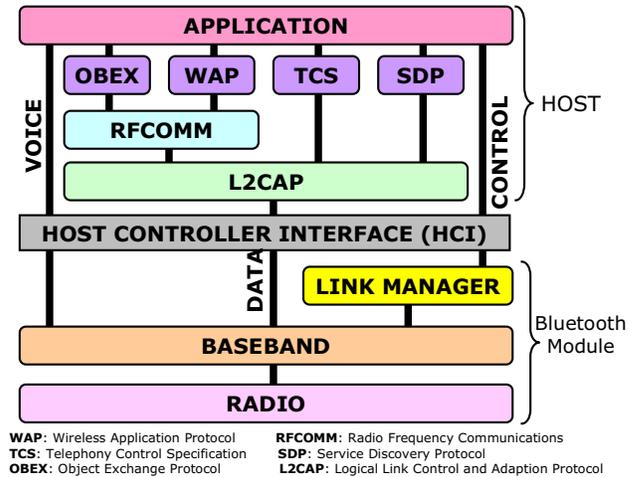

**Fig. 1.** The Bluetooth stack

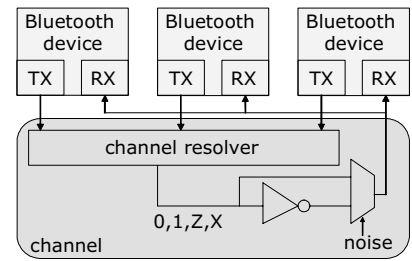

**Fig. 2.** Model of the channel implemented in SystemC

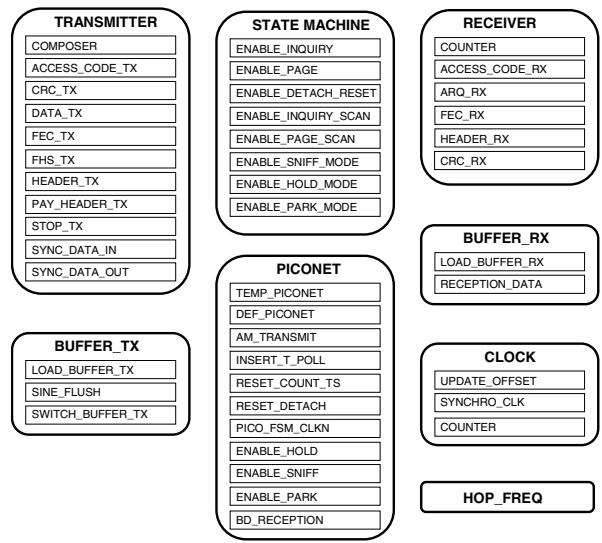

**Fig. 3.** Baseband Architecture implemented in SystemC

## 3. Simulation results

### 3.1. The creation of a PICONET

A Bluetooth piconet is created during the *inquiry* and *page* phases. The main state diagram is reported in Fig. 4.

A device that wants to find other Bluetooth devices enters in the *inquiry* state. In this state the device transmits continuously an inquiry message (an ID type packet) at different frequencies using a frequency hopping sequence defined by a general inquiry access code (GIAC), common to all Bluetooth devices. The GIAC can be used to discover which other Bluetooth units are in range. The dedicated inquiry access code (DIAC) is common for a dedicated group of Bluetooth units that share a common characteristic. The DIAC can be used to discover only these dedicated Bluetooth units. The device that wants to be discovered by other devices enters in the *inquiry scan* state activating the receiver module and waiting for the ID packet from a device in inquiry state. After receiving this ID packet the device enters in the *inquiry response* state in which transmits information that identify itself.

After the inquiry phase, the page phase creates the piconet. The device that wants to become a master enters in the *page* state. In this state the device transmits continuously a page message: an ID type packet containing the BD_ADDR a of the slave he wants to connect in the piconet. The BD_ADDR is a code identifying uniquely each Bluetooth device.

The device that wants to be connected as slave in a piconet enters in the *page scan* state activating the receiver module and waiting for the ID packet with its BD_ADDR from the master device in page state.

Then the master enters in the *master response* state and the slave in the *slave response* state in which the master sends the information to enable the slave to be synchronized to the packets of the piconet.

The SystemC modules of the Link Manager Layer and Baseband Layer have been used to simulate the complex phases of the creation of a piconet. All the signals and the packets can be verified during simulation.

The waveforms related to a simulation of the creation of a piconet with a master and 3 slaves are reported in Fig. 5.

In the simulation reported it has been supposed that all the devices try to connect at the same time, in order to show in a single figure the creation of the complete piconet.

Fig. 5 shows the signals enable_rx_RF sent by the Link Control module of the Baseband Layer to the RF module. The Bluetooth protocol enables the RF transmitter and receiver to switch on only when necessary. This is possible since the master and the slaves are synchronized and they know when a packet can be sent.

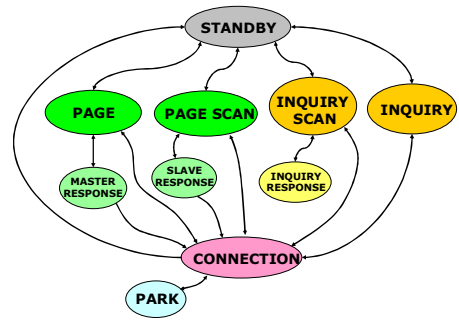

**Fig. 4.** Main state diagram of a Bluetooth device

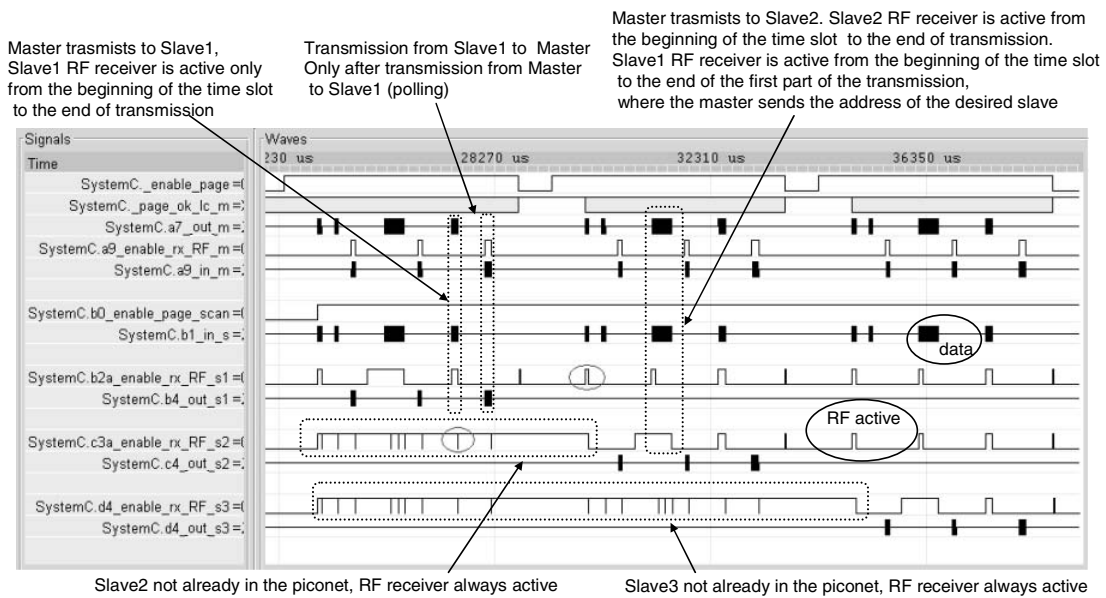

**Fig. 5.** Waveforms for the creation of a piconet with a master and 3 slaves.

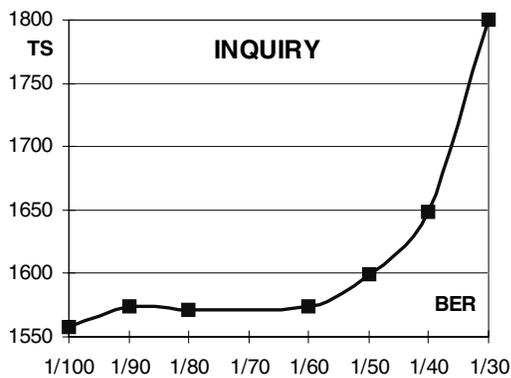

**Fig. 6.** Mean value of time slots required to complete the inquiry phase as a function of BER

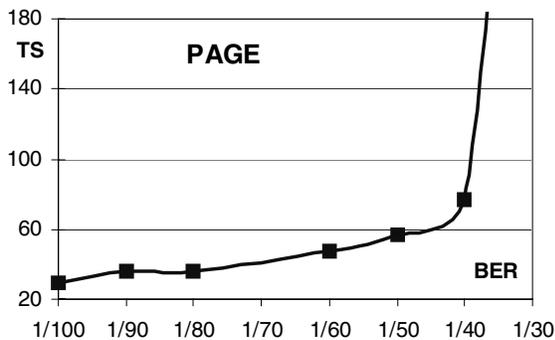

**Fig. 7.** Mean value of time slots required to complete the page phase as a function of BER

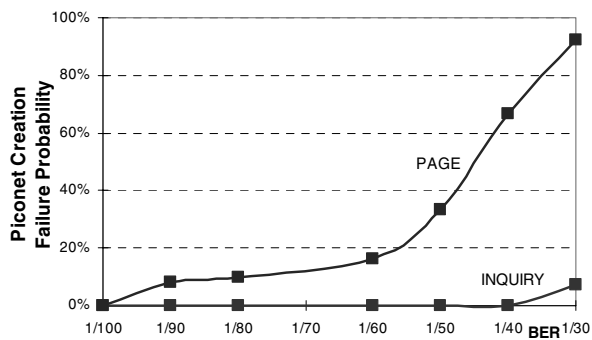

**Fig. 8.** Probability of failure in the creation of a piconet

In Fig. 5 it can be seen that the slaves in page scan, not already in the piconet and therefore not synchronized, have the RF receiver always active. Conversely the enable_rx_RF signal is high only in the beginning of the time slots as soon as they become part of the piconet.

We tested our Bluetooth SystemC implementation in many different situations during the creation of a piconet for a point to point connection and the connection of a piconet up to 4 devices. The executable code is useful to test the possible problems that can be found during the creation of the piconet in order to define the correct algorithms that must be implemented in the application layer. For example the effect of the presence of noise in the channel during the creation of the piconet has been verified. Fig. 6-7 report the mean value of the number of time slots required to complete the inquiry phase and the page phase as a function of the Bit Error Rate (BER), respectively.

During the inquiry phase a specific frequency hopping algorithm is used in the Bluetooth standard [1], in order to increase the speed of connection. The time required to conclude the inquiry phase is random due to the fact that the devices are not already synchronized: the average time is 1556 time slots (TS) in absence of noise in the channel.

The inquiry packets (ID packets of 68 bits) are the less sensitive to the presence of errors in the channel, and almost all the errors are recovered for a BER less than 1/100 without retransmissions.

Only 17 time slots are required to conclude the page phase in absence of noise, since the devices are already synchronized, but the page phase is more sensitive to noise. The correct conclusion of the page phase becomes impossible for a BER higher than 1/30.

The standard defines that the application layer must define a timeout for the inquiry and page phases: the ranges are from 1.28 (2048 TS) to 61.44 secs for the inquiry and from 625 $\mu$secs (1 TS) to 40.9 secs for the inquiry. In the simulations these timeout have been fixed to 1.28 secs for both inquiry and page. The simulations showed that in many cases the piconet is not created for high values of the BER. Fig. 8 shows the percentage of simulations in which the inquiry and the page phases are successfully completed: the probability of success in the page phase is very low with a BER higher than 1/50. Both inquiry and page procedures must be successfully completed to create the piconet: the bottleneck is therefore the page phase.

The statistical results, reported in Figs 6-8, have been obtained with a 0.48 seconds simulation, corresponding of about 775 time slots. The CPU time required was 10'47'' with a speed of 747 clock cycles/secs.

### 3.2. Active, Sniff and hold mode

The Bluetooth standard is a complex standard with many degrees of freedom the designer can use at the application layer in order to reduce the power dissipation, for example: the choice of the different type of packets (DH1, DH3, DH5, DM1, DM3, DM5) and the choice of different modes in which each slave can operate (active, sniff, park and hold).

The Link Controller of the Baseband Layer implements the procedures with the aim of reducing power dissipation. The slaves of the piconet can enter from active mode to sniff, hold or park mode

**Sniff** : the slave does not participate to the piconet, but it becomes active periodically (with a period Tsniff), in order to be synchronized with the piconet itself. During this period of inactivity it cannot participate to other piconets. When it is inactive the transmitter and receiver part of the RF modules are off, with a consistent power reduction.

**Hold**: the slave suspends the connection with the piconet for a predefined amount of time. During this period it can reduce its RF activity or it can participate to another piconet. Before connecting again with the piconet, it must resynchronize to it.

**Park**: the slave remains synchronized with the piconet but it does not take part of the piconet itself. During this period it can reduce its RF activity or it can participate to another piconet. In this way it is possible to insert other slaves in the piconet.

As an example of the simulations performed, Fig. 9 reports the waveforms for a simulation in which two slaves (slave 2 and 3 in the figure) are posed in sniff mode for a sniff_timeout_time of 2 time slots. This can be seen looking at the enable_rx_RF signals in the waveforms.

Many simulations have been performed to derive the RF activity in different mode of functioning of the slave: active, sniff, and hold.

Fig. 10 show the RF activity of the master (transmission and reception) as a function of the frequency of use of the channel, the duty cycle. The duty cycle is defined as the number time slots used for transmission with respect to the maximum time slots available for transmission. The master cannot go in sniff mode, but it does not transmits if it does not need it. The master activate the its reception RF part only in the time slot consequent its transmission since a polling scheme is used in the Bluetooth standard.

Therefore the RF activity of the master is the same if the slave is in sniff or active mode.

Fig. 11 shows the RF activity of the slave as a function of Thold/TS in case it is in active mode or sniff mode. To compare the results between active and sniff mode, it has been supposed that the master transmits data to the slave with a fixed frequency of 100 time slots. A 30% power reduction is obtained with the sniff mode compared to the active mode with Tsniff of 100 time slots, the maximum length for which the slave does not loose packets.

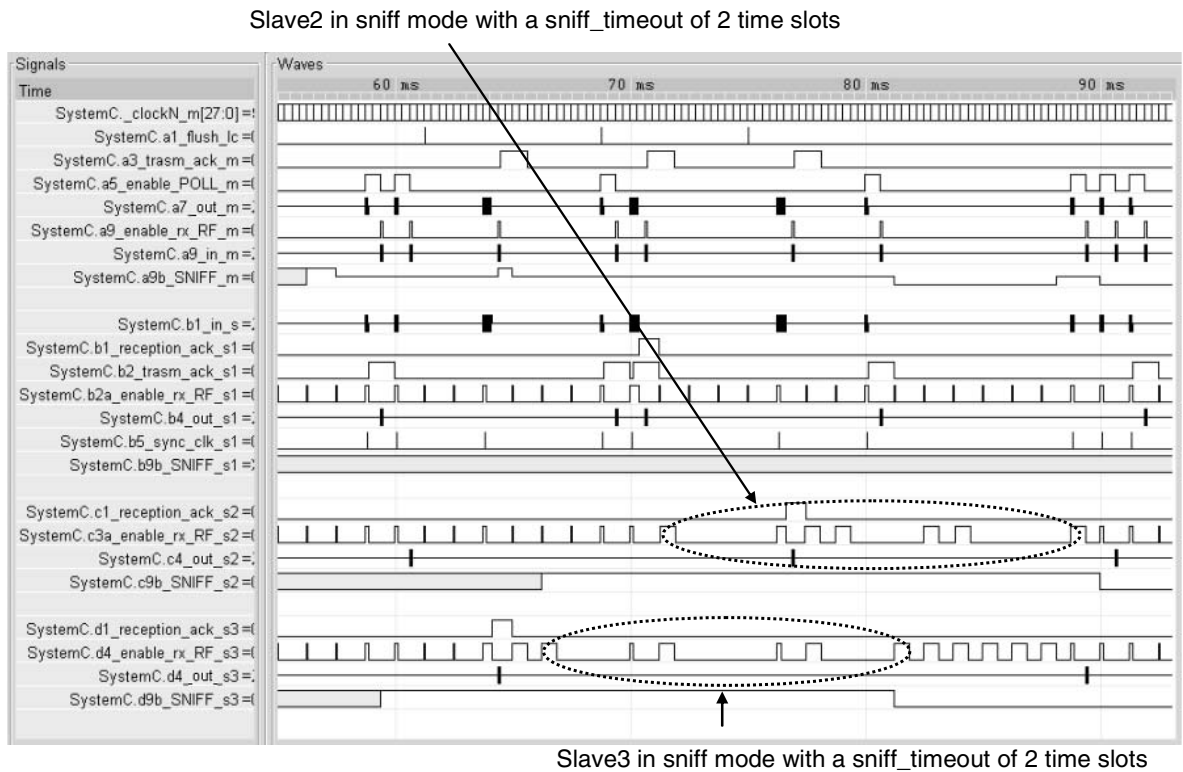

**Fig. 9.** Slave 2 and Slave3 are posed in sniff mode.

Conversely the sniff mode does not allow a reduction in power consumption, if Tsniff is less than 30 time slots.

If the master does not transmits periodically data to the slave, the slave can stay in active mode or in hold mode.

Fig. 12 shows the RF activity of the slave in active or in hold mode as a function of the number of time slots the slave is disconnected to the piconet (Thold/TS).

In active mode the slave RF activity is reduced to the reception of the synchronization packets sent by the master and to a small part of time at the beginning of each time slot to see if the master is transmitting, that is the constant value of 2.6% in Fig. 12.

In hold mode the slave RF activity decreases as the number of time slots the slave is disconnected to the piconet (Thold/TS) increases.

The hold mode gives advantages in power reduction with respect to the active mode only if Thold is higher than 120 time slots.

## 4. Conclusions

The SystemC modules of the Link Manager Layer and Baseband Layer have been used to identify the noise level at which the probability of the creation of a piconet becomes low. The simulations give precise indications to the designer on the power that can be reduced using sniff and hold mode in the different situations of use of the Bluetooth network.

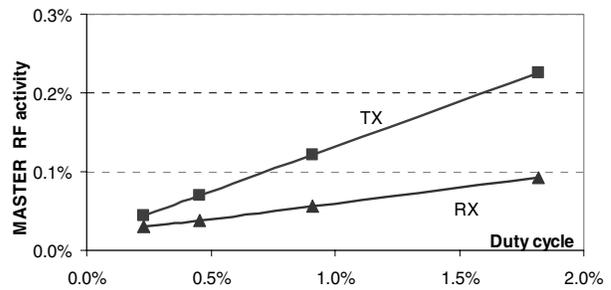

**Fig. 10.** RF activity of the master as a function of the frequency of use of the channel

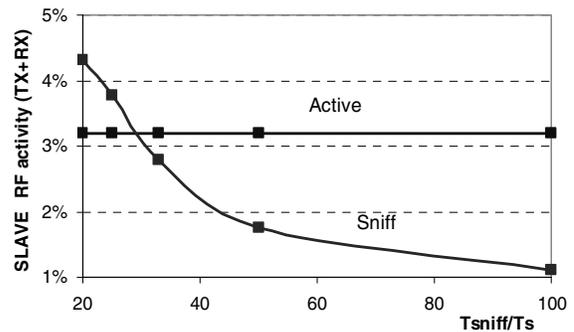

**Fig. 11.** RF activity of the slave as a function of Tsniff

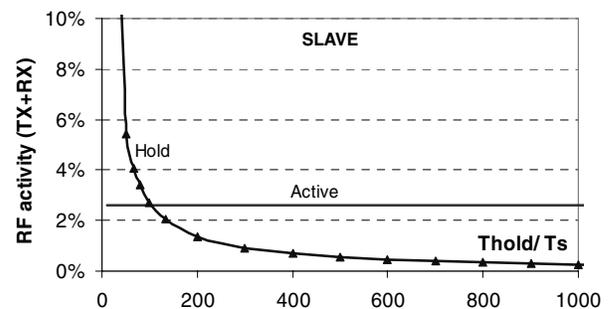

**Fig. 12.** RF activity of the slave as a function of the hold time.